\documentclass{blois}

\usepackage[inter-unit-product =\cdot]{siunitx}
\def\Journal#1#2#3#4{{#1} {\bf #2}, #3 (#4)}

\def\PRL{\em Phys. Rev. Lett.}

\def\PRD{{\em Phys. Rev.} D}

\def\be{\begin{equation}}
\def\ee{\end{equation}}
\def\bea{\begin{eqnarray}}
\def\eea{\end{eqnarray}}

\newcommand{\STEREO}{\textsc{Stereo} }
\newcommand{\LLR}{$Q_{tail}$/$Q_{tot}$ }
\newcommand{\Photo}{}

\begin{document}
\vspace*{4cm}
\title{Search for eV Sterile Neutrinos -- The STEREO Experiment [Blois 2019]}

\author{S. Schoppmann \footnote{for the \STEREO collaboration}}

\address{Max-Planck-Institut f\"ur Kernphysik, Saupfercheckweg 1, 69117 Heidelberg, Germany}

\maketitle\abstracts{
The \STEREO experiment is designed to test the hypothesis of light sterile neutrinos being the cause of the Reactor Antineutrino Anomaly. It measures the antineutrino energy spectrum from the compact core of the ILL research reactor in six identical detector cells covering baselines between 9 and 11\,m. Results from 119~days of reactor turned on and 211~days of reactor turned off are reported. Using a direct comparison between neutrino interaction rates of all cells, independent of any flux prediction, we find compatibility with the null oscillation hypothesis. The best fit point of the Reactor Antineutrino Anomaly is rejected at 99\,\% C.L. }

\section{\label{sec:intro}Introduction}
In 2011, a re-evaluation of the prediction of anti-neutrino spectra emitted by nuclear reactor cores lead to a 6.5\,\% deficit between detected and expected fluxes at baselines shorter than 100\,m \cite{reactorSpectrum1}. The deficit, known as reactor anti-neutrino anomaly (RAA) could either be caused by underestimated systematics in the prediction steps of the neutrino flux or by a beyond-standard-model sterile, i.e.~non-weakly interacting, additional neutrino. Its existence would manifest in a baseline and energy dependent distortion of the measured energy spectra. Thus, by segmenting the \STEREO detector, a prediction-independent relative measurement can be performed. A maximal oscillation effect is expected below 10\,m~\cite{reactorSpectrum1}.

Among \STEREO\cite{stereoPhase1}, a series of other experiments conduct measurements at these baselines~\cite{reviewBuck}. All report significant exclusions of the allowed parameter space~\cite{reactorSpectrum1}. However, the combination of result from DANSS, and NEOS in a global shape-only analysis suggests the existence of a sterile neutrino in contrast to findings by NEUTRINO-4 which imply another sterile oscillation~\cite{reviewBuck}. How this tension will be resolved is yet unclear. Some results await final calculation of systematic uncertainties or rely on non-trivial prediction-dependent conversion between different reactor and detector types.

\section{Experiment Description}
\begin{figure}[tb]
    \begin{center}
	\includegraphics[width=0.5\linewidth]{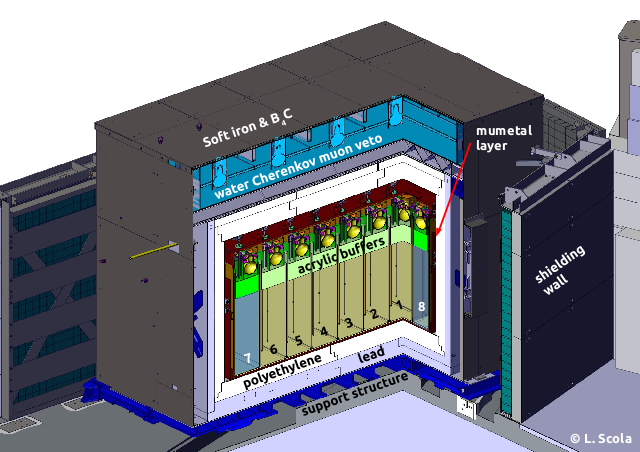}
    \end{center}
    \caption{\STEREO detector setup. 1--6: Target cells (baselines from reactor core: 9.4 -- 11.1\,m); 7, 8: two of the four Gamma-Catcher cells.}
    \label{fig:detector}
\end{figure}

The \STEREO detector~\cite{stereoDetector} is located at the ILL research centre located in Grenoble, France, at 15~m.w.e.~overburden. Its 93\,\% enriched ${}^{235}\textrm{U}$ reactor is a compact heavy water reactor of 58.3\,MW nominal thermal power. The \STEREO detector measures neutrinos via an inverse beta decay (IBD) reaction in gadolinium (Gd) loaded organic liquid scintillator~\cite{stereoScintillator}: $\bar{\nu}_{e} + p \rightarrow e^{+} + n$. The positron yields a prompt annihilation signal, while the neutron gives a delayed capture signal after it has been captured by a Gd isotope. The fiducial target volume (TG) of 2.4\,m${}^{3}$ is divided into six identical and optically separated cells towards the direction of the neutrinos (cf. Figure \ref{fig:detector}). This allows six flux measurements at baselines between 9.4 and 11.1\,m.

The energy scale of the \STEREO detector is closely followed over time by regular calibrations. Sources can be deployed inside each cell as well as around and underneath the inner detector part. Each system allows deployment of various, partly custom-made, gamma and neutron sources spanning gamma energies between 0.5 and 4.4\,MeV~\cite{stereoDetector}.

Particular emphasis in the simulation of \STEREO was given to the simulation of gammas released by the capture of a neutron by Gd. New measurements and calculations of the energy levels of the relevant Gd isotopes ${}^{155}\textrm{Gd}$ and ${}^{157}\textrm{Gd}$ were introduced into the simulation exploiting the FIFRELIN package \cite{FIFRELIN}. It includes latest experimental nuclear data. An improved agreement between data and simulation spectra was found throughout the full energy range. Further details can be found in \cite{stereoFIFRELIN,stereoFIFRELINdata}.

\section{\label{sec:result}Oscillation Analysis}
During its operational time, \STEREO recorded two datasets denoted as ``phase-I'' and ``phase-II'' with a long reactor shut-down in 2017 separating both datasets~\cite{STdata}. Phase-II spans 199~days of reactor-on and 211~days of reactor-off data with a total of 43400 neutrino candidate events. Reactor-on and short reactor-off periods alternate, allowing for measurements of background-pure samples roughly about every two months. In this article, only the new data phase-II is considered, as the combination of phases is currently in progress.

The first step of background discrimination is done in a cut-based fashion. A summary of all cuts is given in Table \ref{tab:selectionCuts}.
\begin{table}[tb]
    \caption[]{Selection cuts for IBD candidates.}
    \label{tab:selectionCuts}
    \vspace{0.4cm}
    \begin{center}
    \begin{tabular}{|l|r|l|r|l|}
    \hline
    Type        &\#& Requirement for passing cut &\# & Requirement for passing cut\\
    \hline
    Energy      &1& $\SI{1.625}{MeV} < E^\textrm{detector}_\textrm{prompt} < \SI{7.125}{MeV}$ &2& $\SI{4.5}{MeV} < E^\textrm{detector}_\textrm{delayed} < \SI{10}{MeV}$ \\
    \hline
    Coincidence &3& $\SI{2}{\micro\second} < \Delta T_\textrm{prompt-delayed} < \SI{70}{\micro\second}$ &4& $\Delta X_\textrm{prompt-delayed} < \SI{600}{\milli\metre}$ \\
    \hline
    Topology    &5& $E^\textrm{cell}_\textrm{prompt} < \SI{1}{MeV}$, neighbour cells &7& $E^\textrm{Target}_\textrm{delayed} > \SI{1}{MeV}$ \\
                &6& $E^\textrm{cell}_\textrm{prompt} < \SI{0.4}{MeV}$, other cells & & \\
                
    \hline
    Muon-       &8& $\Delta T^\textrm{veto}_\textrm{muon-prompt} > \SI{100}{\micro\second}$ &10& $\Delta T_\textrm{event-prompt} > \SI{100}{\micro\second}$ and \\ 
    induced     &9& $\Delta T^\textrm{detector}_\textrm{muon-prompt} > \SI{200}{\micro\second}$ & & $\Delta T_\textrm{delayed-event} > \SI{100}{\micro\second}$ \\ 
    background

                &11& $\frac{Q_\textrm{PMT max, prompt}}{Q_\textrm{cell, prompt}} < 0.5$ & & if $E^\textrm{detector}_\textrm{event} > \SI{1.5}{MeV}$ \\
    \hline
    \end{tabular}
    \end{center}
\end{table}
Cuts \#1 to \#4 aim for the characteristic energy and correlation between the two sub-events of the IBD. Cuts \#5 to \#7 ensure that the event can clearly be attributed to a TG cell. Cuts \#8 to \#11 reject muons and events induced by muons, e.g.~multiple fast neutron cascades, decaying muons, and others. The overall acceptance in the TG is 61\,\% showing a slight correlation with energy of 4\,\% between 3 and 8\,MeV.

After the cut-based selection, the remaining background is further discriminated by exploiting the time shape of the selected prompt scintillator pulses. As parameter for this pulse shape discrimination (PSD), the late light ratio \LLR (LLR), i.e. the ratio between the charge in the tail of a scintillation pulse and the total charge of the pulse, is chosen. Two populations can be distinguished in terms of \LLR which are made up from IBD events, correlated electronic background induced by cosmic rays, and accidental coincidences (single rates are dominated by gammas) on the one hand and muon-induced fast neutrons one the other hand.

To extract the number of IBD candidates from the distribution, the background distribution in the region of interest needs to be know. This distribution can be estimated in-situ from the reactor-off phases. However, it is not possible to apply the background spectra directly as they scale, e.g.~with temperature or atmospheric pressure. Thus, spectra are corrected for those effects exploiting different temperature and atmospheric pressure settings during reactor-off phases. Furthermore, this technique allows us to verify a stable shape of the LLR distributions across data taking periods.

The oscillation analysis is performed in 13 equidistant energy bins of 500\,keV width between 1.625\,MeV and 7.125\,MeV. The number of IBD-candidates for each of the six cells is relatively compared against a common expectation via a $\Delta \chi^2$ method with
\begin{equation}
    \chi^2 = \sum_{l=1}^{N_\textrm{cells}} \sum_{i=1}^{N_\textrm{Ebins}} \left(\frac{D_{l,i} - \phi_i M_{l,i}}{\sigma_{l,i}}\right)^2 
    +\sum_{l=1}^{N_\textrm{cells}}\left(\frac{\alpha_{l}^{\textrm{EscaleU}}}{\sigma_{l}^{\textrm{EscaleU}}}\right)^2+\left(\frac{\alpha^{\textrm{EscaleC}}}{\sigma^{\textrm{EscaleC}}}\right)^2 + \sum_{l=1}^{N_\textrm{cells}}\left(\frac{\alpha_{l}^{\textrm{NormU}}}{\sigma{l}^{\textrm{NormU}}}\right)^2
\end{equation}
being used. With $l$ and $i$ being indices running over all cells and energy bins, respectively, $D_{l,i}$ and $\sigma_{l,i}$ denote the numbers of measured neutrino candidates and their statistical uncertainties. The $M_{l,i}$ are the corresponding expected numbers of neutrinos. They depend on the oscillation parameters and set of nuisance parameters $\vec{\alpha}$.

Since we do not want to rely on absolute rate predictions in our analysis, the $\phi_{i}$ are introduced as free normalisation parameters. For each energy bin $i$, they effectively adjust the number of expected neutrinos $M_{l,i}$ across all cells $l$ to match the number of measured neutrino candidates $D_{l,i}$ on average. The $M_{l,i}$ are then optimised in terms of oscillation parameters $\sin^2(2\theta_{ee}), \Delta m_{41}^2$ and nuisance parameters $\vec{\alpha}$ to match the remaining deviations from the $D_{l,i}$ in each cell $l$. Since the $\phi_{i}$ absorb all absolute rate information, the analysis becomes independent of the spectrum prediction. Moreover, cell-to-cell correlated uncertainties do not affect the result.

The no-oscillation scenario was tested by comparing the $\Delta \chi^2$ of data with the distribution from pseudo-experiments. We get a p-value of 0.4, i.e.~the no-oscillation hypothesis cannot be rejected. In order to reject allowed parameter space points of the RAA, the parameter space is scanned in a raster of variable but fixed $\Delta m_{41}^2$ values.
\begin{figure}[tb]
    \begin{center}
	\includegraphics[width=0.5\linewidth]{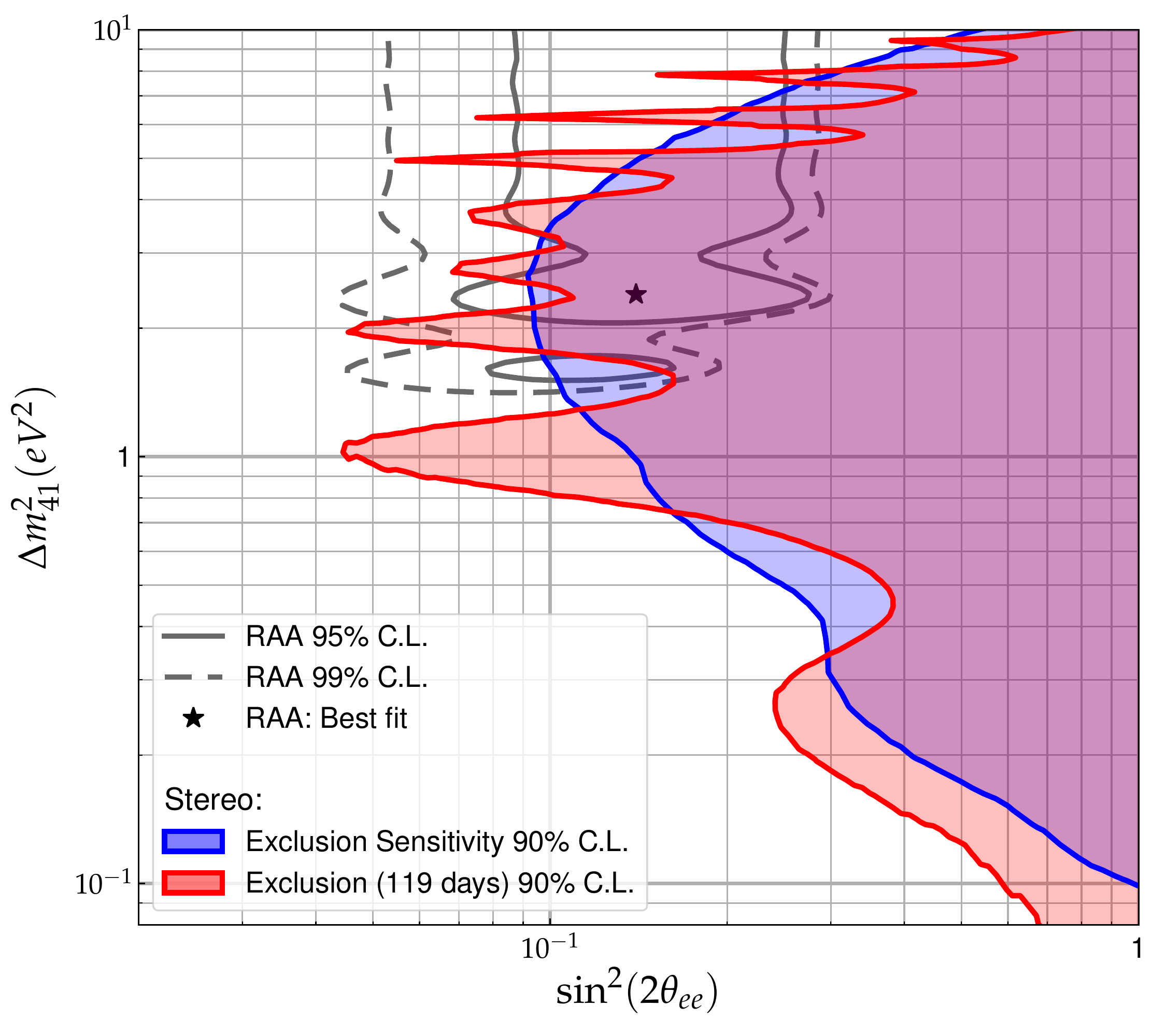}
	\end{center}
	\caption{Rejection contour of phase-II at 90\,\% C.L.~(red) compared to the expected sensitivity curve (blue). The statistical fluctuations of the rejection contour are located around the sensitivity contour, as expected. Overlaid are the allowed regions of the RAA (grey) and their best-fit point (star).}
	\label{fig:contour}
\end{figure}
By plotting the rejected intervals of $\sin^2(2\theta_{ee})$ for each $\Delta m_{41}^2$ value along the ordinate, we achieve the exclusion region depicted in Figure~\ref{fig:contour}. In particular, we reject the best fit point of the RAA at the 99\,\% C.L.~for its corresponding $\Delta m_{41}^2$ value. The results are confirmed with high agreement by an independent $\Delta \chi^2$ method, where all uncertainties are modelled by a covariance matrix instead of nuisance parameters.

\section*{Acknowledgments}
This work is funded by the French National Research Agency (ANR) within the project ANR-13-BS05-0007 and the ``Investments for the future'' programmes P2IO LabEx (ANR-10-LABX-0038) and ENIGMASS LabEx (ANR-11-LABX-0012). We are grateful for the technical and administrative support of the ILL for the installation and operation of the \STEREO detector. We further acknowledge the support of the CEA, the CNRS/IN2P3 and the Max Planck Society.

\end{document}